\documentclass[fleqn,usenatbib]{mnras}

\usepackage{newtxtext,newtxmath}

\usepackage[T1]{fontenc}

\DeclareRobustCommand{\VAN}[3]{#2}
\let\VANthebibliography\thebibliography
\def\thebibliography{\DeclareRobustCommand{\VAN}[3]{##3}\VANthebibliography}

\usepackage{graphicx}	
\usepackage{amsmath}	
\usepackage{subcaption}
\usepackage{hyperref}

\newcommand{\source}{1RXH~J173523.7--354013}
\newcommand{\rx}{RX~J1735.3--3540}
\newcommand{\rxs}{1RXS~J173524.4--353957}
\newcommand{\igr}{IGR~J17353--3539}

\newcommand{\Msun}{\mathrm{M}_{\odot}}
\newcommand{\Rsun}{\mathrm{R}_{\odot}}
\newcommand{\lum}{\mathrm{erg~s}^{-1}}

\newcommand{\xmm}{\textit{XMM-Newton}}

\def \aj {AJ}
\def \mnras {MNRAS}
\def \apj {ApJ}
\def \apjs {ApJS}
\def \apjl {ApJL}
\def \aap {A\&A}
\def \nat {Nature}

\def\farcs{\hbox{$.\!\!^{\prime\prime}$}}

\title[The nature of the VFXB 1RXH J173523.7--354013]{The nature of very-faint X-ray binaries: Near-infrared spectroscopy of 1RXH J173523.7--354013 reveals a giant companion}

\author[Shaw et al.]
{A. W. Shaw,$^{1,2}$\thanks{E-mail: awshaw@butler.edu} 
N. Degenaar,$^{3}$\thanks{E-mail: degenaar@uva.nl} 
T. J. Maccarone,$^{4}$ 
C. O. Heinke,$^{5}$ 
R. Wijnands,$^{3}$ 
J. van den Eijnden$^{6}$
\\
$^{1}$Department of Physics and Astronomy, Butler University, 4600 Sunset Ave, Indianapolis, IN, 46208, USA\\
$^{2}$Department of Physics, University of Nevada, Reno, 1664 N. Virginia Street, Reno, NV, 89557, USA\\
$^{3}$Anton Pannekoek Institute for Astronomy, University of Amsterdam, Science Park 904, 1098 XH, Amsterdam, the Netherlands\\
$^{4}$Department of Physics \& Astronomy, Texas Tech University, Box 41051, Lubbock, TX, 79409-1051, USA\\
$^{5}$ Department of Physics, University of Alberta, CCIS 4-181, Edmonton, AB, T6G 2E1, Canada\\
$^{6}$Department of Physics, University of Warwick, Coventry CV4 7AL, UK
}

\date{Accepted 2023 November 22. Received 2023 November 21; in original form 2023 September 30}

\pubyear{2021}

\begin{document}
\label{firstpage}
\pagerange{\pageref{firstpage}--\pageref{lastpage}}
\maketitle

\begin{abstract}
Very-faint X-ray binaries (VFXBs) are a sub-class of black holes and neutron stars in binaries that appear to be accreting at a very low rate. In addition to providing interesting constraints on poorly understood forms of accretion, elucidating the nature of VFXBs is particularly interesting for binary evolution and population modeling. Through near-infrared (nIR) spectroscopy, we here investigate the nature of the bursting neutron star and VFXB 1RXH J173523.7--354013 (J1735), which persistently accretes at an X-ray luminosity of $L_X \sim 10^{34} - 10^{35}~\lum$. Our analysis shows that the nIR emission is dominated by that of the companion star, which we find to be a late G or early K-type giant, making this the second neutron star identified as a VFXB found to have a giant companion. We discuss how several of the system properties are difficult to reconcile with a wind-fed symbiotic X-ray binary. We therefore also propose an alternative scenario wherein J1735 is a wide binary system (supported by the discovery of a $7.5$ d modulation in the nIR light curves) with a quiescent luminosity of $L_X \sim 10^{34} - 10^{35}~\lum$, in which the donor star is overflowing its Roche lobe. This raises the possibility that J1735 may, every century or more, exhibit very long and very bright outbursts during which it reaches accretion rates around the Eddington limit like the neutron star Z sources.
\end{abstract}


\section{Introduction}
Among the brightest X-ray point sources in the sky are neutron stars (NSs) and black holes (BHs) that are in a binary system accreting gas from a less massive companion star. These low-mass X-ray binaries (LMXBs) can generate an X-ray luminosity of $L_X \simeq 10^{36-39}~\lum$ when actively accreting. However, a significant sub-set of the LMXBs \citep[comprising $\simeq$25\% of the currently known population;][]{Bahramian2022} has a much lower peak accretion output of $L_X \simeq 10^{34-36}~\lum$: the very faint X-ray binaries \citep[VFXBs;][]{Wijnands2006}. 

Due to their low luminosity, detailed studies of VFXBs were long hampered by sensitivity limitations of astronomical instrumentation. Whereas there has been a significant uprise in targeted studies to find and characterize VFXBs over the past $\simeq$2 decades \cite[e.g.,][]{Sakano2005,Muno2005,intZand2005a,Wijnands2006,wijnands2008,Degenaar2010,ArmasPadilla2011,Sidoli2011,Shaw2020,Bahramian2021}, many aspects of the accretion morphology and binary configurations of this large LMXB sub-group remain to be explored. There is significant interest in this, because VFXBs are a gateway to study accretion processes and outflows in an under-explored mass-accretion regime \cite[e.g.,][]{ArmasPadilla2013a,Knevitt2014,Wijnands2015,degenaar2017,Gusinskaia2020}. Moreover, elucidating the binary configuration of VFXBs allows to test and refine various binary evolution and population synthesis models \cite[e.g.,][]{Pfahl2002b,King2006,Yungelson2006,vanHaaften2012,Maccarone2013,Maccarone2015}. 

A natural explanation for the X-ray faintness of VFXBs may be that these LMXBs have small accretion disks \citep[i.e. short orbital periods; e.g.][]{King2006,intZand2007,Heinke2015}. This includes the sub-population of ultra-compact X-ray binaries (UCXBs), which have degenerate donor stars and hence require orbital periods $\lesssim$90~min to facilitate accretion through Roche-lobe overflow \citep[e.g.][]{Nelson1986}. Indeed, by now several VFXBs have been conclusively identified as UCXBs \citep[see][and references therein]{Bahramian2022}. Nevertheless, at least three VFXBs, \source\ (aka \rxs, \rx, \igr; J1735 hereafter), M15 X-3, and SAX\,J1806.5$-$2215, must have non-degenerate donor stars; strong H$\alpha$ emission was detected in the optical spectrum of J1735 \citep[][]{Degenaar2010burst}, the high optical/X-ray ratio of M15 X-3 is inconsistent with an UCXB interpretation \cite[][]{Arnason2015}, and SAX\,J1806.5$-$2215 appears to exhibit absorption in Brackett $\gamma$, providing evidence for hydrogen in the system and therefore excluding a hydrogen-stripped companion \citep{Shaw2017a}. 

Another possible explanation for the faint X-ray emission of VFXBs is that the accretion flow is radiatively-inefficient. Standard accretion theory predicts that towards low accretion rates the radiative efficiency decreases \citep[][]{rees1982,narayan1995}. 
This should be particularly pronounced for BH systems with short (though not ultra-compact) orbital periods \citep[][]{Wu2010,Knevitt2014}, of which indeed a handful of candidates have been identified among the VFXBs \citep[][]{McClintock2001a,CorralSantana2013,Stoop2021}. In NS systems, the strong NS magnetic field may also hamper efficient accretion onto the stellar surface \citep[e.g.,][]{wijnands2008,degenaar2014,Heinke2015}. Several VFXBs have been identified as accreting millisecond pulsars \citep[e.g.,][]{Altamirano2010,Strohmayer2017,Sanna2018,Sanna2022}. For one of these the inner accretion disk was found to be truncated and hints of outflowing material were found, suggesting that some material might be propelled out \cite[][]{degenaar2017,vandeneijnden2018}. 

Finally, recent systematic surveys to find and identify VFXBs have revealed that symbiotic systems \citep[e.g.][]{Iben1996,Corbet2008,Bozzo2022} could be a growing sub-class \citep[][]{Shaw2020,Bahramian2021}. In these systems the compact object (which can be a NS or even a white dwarf) accretes from the weak wind of the low-mass companion, which is inefficient, and hence can explain the low X-ray luminosity. A somewhat surprising aspect of this is that the one VFXB identified as a symbiotic X-ray binary \citep[IGR\,J17445$-$2747;][]{Shaw2020,Bahramian2021} displays thermonuclear bursts \citep[][]{Mereminskiy2017a}. These bursts \citep[see e.g.,][for a review]{Galloway2021} are often regarded as exclusively occurring 
in LMXBs, since bursts are suppressed or have a different observational signature on high $B$ field NSs 
\citep[][]{Bildsten1997}, 
and most low-$B$ NSs are found in LMXBs. 
Symbiotic systems, on the other hand, are a short-lived stage of binary evolution such that there is not enough time for accretion to reduce the magnetic field of the NS to a level where 
bursts can be produced. 
Bursts from a symbiotic system therefore strongly suggest that the NS was born with a low $B$ field. 
In summary, the VFXBs appear to be an inhomogeneous group, consisting of several sub-classes of LMXBs: UCXBs, short-period (though not ultra-compact) BH systems, accreting millisecond X-ray pulsars and symbiotic X-ray binaries. 

In this work, we investigate the nature of the VFXB J1735. \cite{Degenaar2010burst} identified this system as a NS-LMXB based on the detection of a thermonuclear burst. Reprocessed X-ray burst emission was clearly detected at optical and UV wavelengths, which allowed them to pin down a unique counterpart for the X-ray source. Ultraviolet (UV), optical and near-infrared (nIR) photometry indicated that the counterpart was rather red, while optical spectroscopy revealed a strong H$\alpha$ emission line \citep[][]{Degenaar2010burst}. Deep radio observations of the source did not yield a detection \citep[][]{vandeneijnden2021}, which implies that it is unlikely that its bright nIR emission is due to the contribution of a jet. {\em Gaia} Data Release 3 \citep[DR3;][]{GaiaCollaboration2016,GaiaCollaboration2022} measured the parallax to the optical counterpart to be $p=1.82\pm0.56$ mas, and \citet{BailerJones2021} calculated the photogeometric distance to be $d=3.5^{+2.3}_{-1.3}$ kpc based on {\em Gaia} Early Data Release 3 \citep[EDR3;][]{GaiaEDR3summary2021}. In this paper we further investigate the nature of this VFXB using nIR spectroscopy.

\section{Observations and Data Reduction}
\label{sec:observations}

\subsection{Near-infrared monitoring}

\begin{figure}
    \centering
     \begin{subfigure}[h]{0.475\textwidth}
         \centering
         \includegraphics[width=\textwidth]{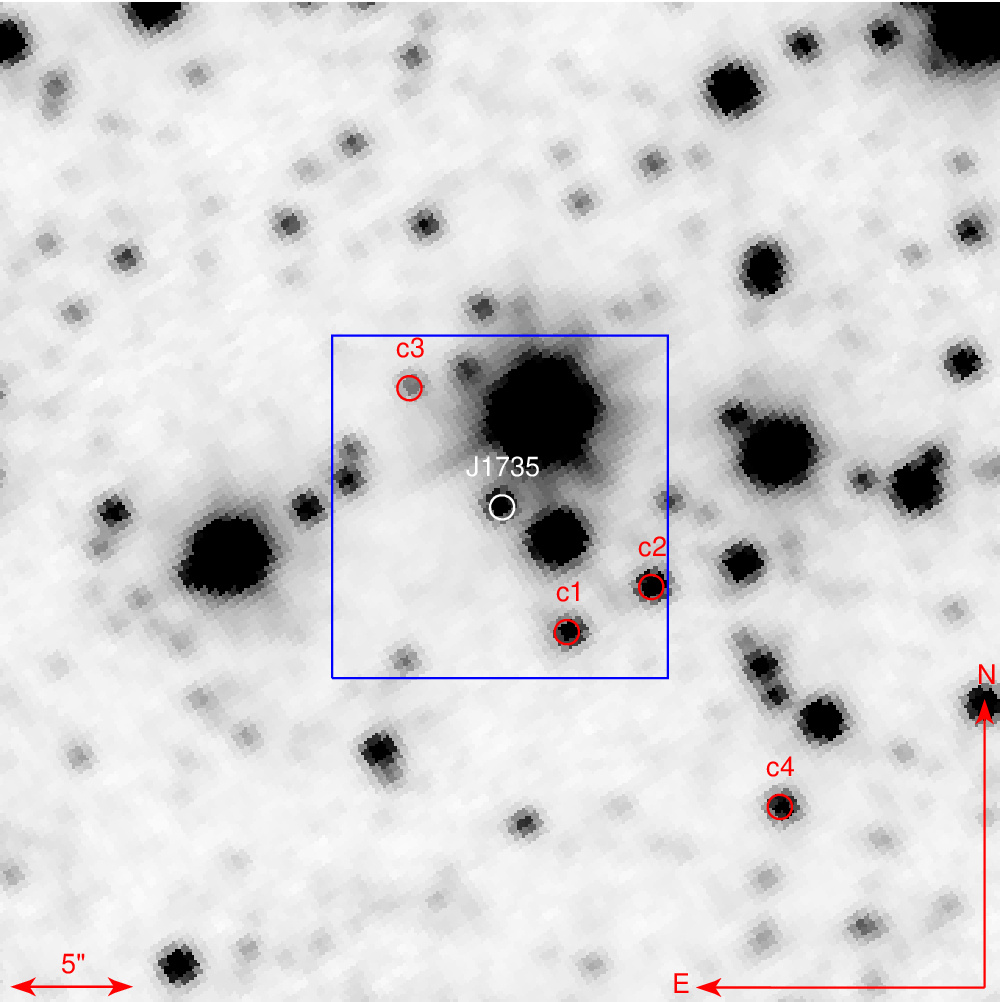}
     \end{subfigure}
     \newline
     \begin{subfigure}[h]{0.475\textwidth}
        \centering
        \includegraphics[width=\textwidth]{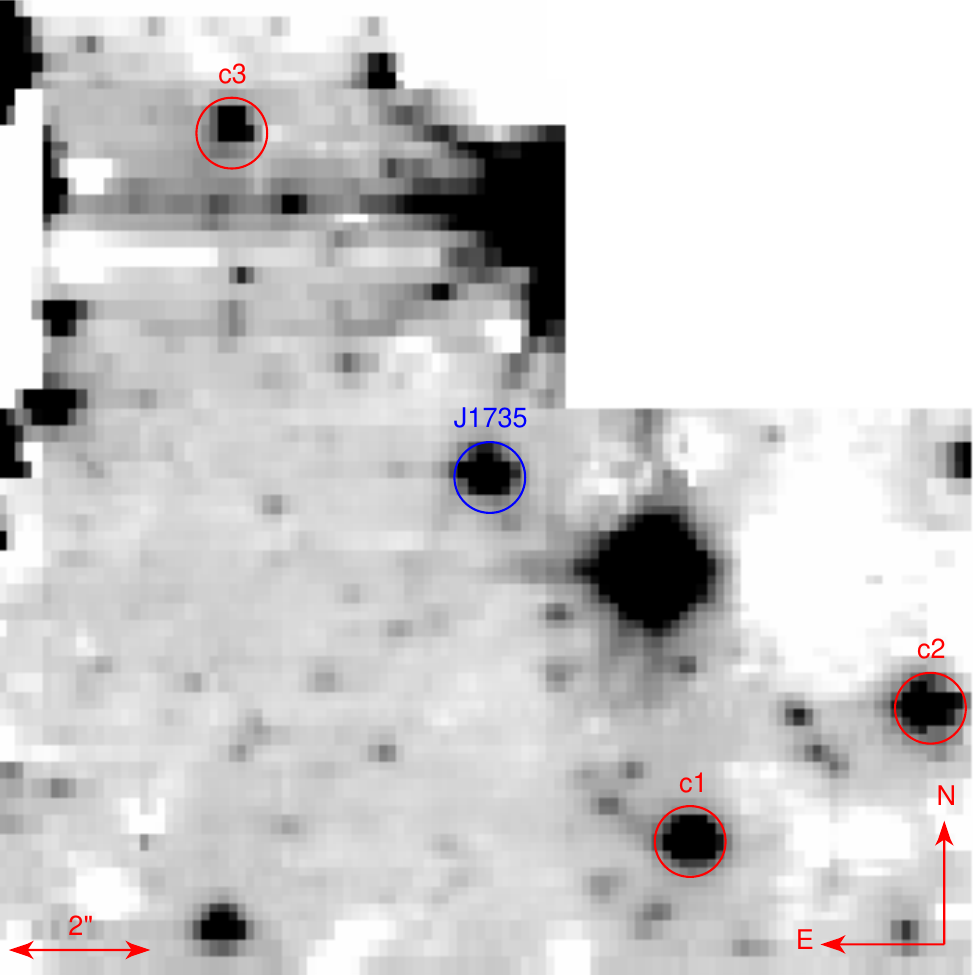}
     \end{subfigure}
     \caption{Images of the J1735 field. {\it Upper panel:} VVV $K_s$-band image of the field, with red circles at the positions of the comparison stars we use to construct the differential light curves in Fig. \ref{fig:vvv_lc}. The white circle is at the position of the nIR counterpart to J1735, as identified by \citet{Degenaar2010burst}. The blue box represents the approximate FOV of the VLT/SINFONI spectroscopic follow-up. {\it Lower panel:} SINFONI $H+K$-band median cube image of the J1735 field, with red circles showing some of the same comparison stars from the upper panel. The nIR counterpart to J1735 is highlighted in this panel by a blue circle. White patches in the lower panel represent artefacts from the dithering pattern of SINFONI and the subsequent sky subtraction, but this does not affect the extracted spectrum of J1735.}
    \label{fig:VVV_SINFONI_image}
\end{figure}

\begin{figure*}
    \centering
    \includegraphics[width=0.95\textwidth]{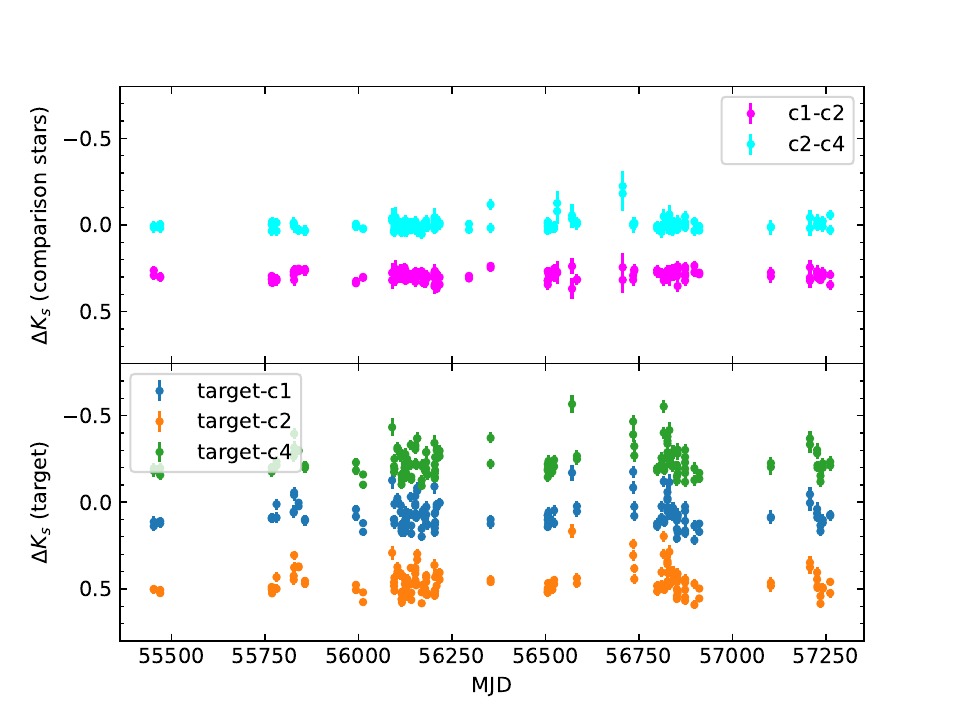}
    \caption{{\it Lower panel:} VVV differential $K_{\rm s}$-band light curves of the nIR counterpart to J1735 using a variety of comparison stars with similar magnitudes (c1, c2 and c4). {\it Upper panel:} VVV differential $K_{\rm s}$-band light curves of the comparison stars, showing minimal variability. In both panels, light curves are offset from each other for clarity.}
    \label{fig:vvv_lc}
\end{figure*}

We queried data release 5 (DR5) of the Visible and Infrared Survey Telescope for Astronomy (VISTA) Variable in the Via Lactea Survey \citep[VVV;][]{Minniti2010} for nIR imaging of the J1735 field. We found that the field had been observed multiple times  since 2010 Sept, with the densest coverage in the $K_s$-band. A $K_s$-band image of the field is shown in the upper panel of Fig. \ref{fig:VVV_SINFONI_image}. We downloaded the source detection table for the nIR counterpart of J1735 (known as VVV\,J173523.74-354016.58), which contains aperture photometry for numerous values of aperture radius.\footnote{For an overview of VISTA science archive data, see \href{http://horus.roe.ac.uk/vsa/dboverview.html}{http://horus.roe.ac.uk/vsa/dboverview.html} }. We also downloaded source detection tables for four comparison stars (see Fig. \ref{fig:VVV_SINFONI_image}). We extracted the 1\arcsec\ radius \citep[as recommended by][for the crowded field case]{Saito2012} aperture photometry from the source detection tables for J1735 and the comparison stars and assembled differential $K_s$-band light curves, plotting them in Fig. \ref{fig:vvv_lc}.\footnote{For clarity, we do not include the third comparison star in Fig. \ref{fig:vvv_lc}, but the observed variability is nearly identical to what is shown.} We choose to calculate differential magnitudes instead of simply plotting the VVV apparent magnitudes, because this allows us to assess whether any variability is intrinsic to the J1735 system, or is an artefact of performing aperture photometry in a crowded field.

\subsection{Near-infrared spectroscopy}

The J1735 field was observed with the Spectrograph for INtegral Field Observations in the Near
Infrared \citep[SINFONI;][]{Eisenhauer2003} on the Very Large Telescope at Paranal, Chile on 2018 August 03. Target of Opportunity (ToO) observations were obtained as part of proposal 0102.D-0650 (PI: Degenaar). We obtained 14 total images, the first ten of which consisted of 5 co-added 20s exposures while each of the remaining 4 images consisted of 4 co-added 20s exposures, all of which were obtained with the $H+K$ grating ($1.45-2.45$ $\mu$m). We utilised the 0\farcs25 spatial pixel scale, corresponding to an 8\arcsec$\times$8\arcsec field-of-view for each image. A dithering pattern was applied to account for the variable nIR sky background. We used a nearby bright star as a natural guide star (NGS) to correct for atmospheric distortions with the SINFONI adaptive optics (AO).

Data were reduced with the ESO Recipe Execution tool ({\sc esorex}) pipeline,\footnote{\href{https://www.eso.org/sci/software/cpl/esorex.html}{https://www.eso.org/sci/software/cpl/esorex.html}} which performs typical reduction steps including dark subtraction, non-linearity correction, flat fielding, wavelength calibration and sky subtraction to provide a co-added data cube, from which the median image in Fig. \ref{fig:VVV_SINFONI_image} was derived. We extracted the spectrum of the nIR counterpart to J1735 with the {\sc esorex} task {\tt sinfo\_utl\_cube2spectrum}, using a circular region of radius 2 pixels (0\farcs5).

We corrected for telluric features in the $H+K$ spectrum of J1735 using the software {\sc molecfit} \citep{Smette2015,Kausch2015}, which uses the source spectrum itself as an input to fit and remove telluric lines. The advantage of {\sc molecfit} over the typical method of observing a telluric standard is that using the science spectrum as an input means the derived telluric spectrum is representative of the nIR sky spectrum under the actual observing conditions. Telluric standards are typically observed at different times, airmasses and sky background conditions and this can affect the accuracy of the telluric correction of the science spectrum. 

\begin{figure}
    \centering
    \includegraphics[width=0.475\textwidth]{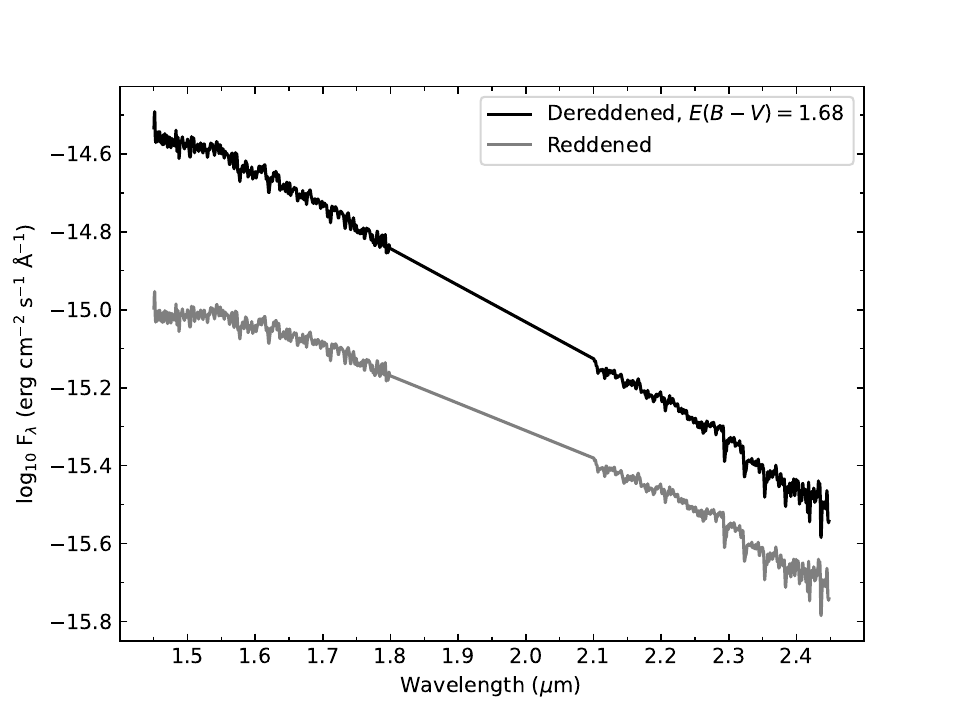}
    \caption{Full $H+K$ spectrum of the nIR counterpart to J1735, flux calibrated. The observed spectrum is shown in grey and the dereddened spectrum in black, which was calculated assuming $E(B-V)=1.68$ \citep{Degenaar2010burst} and a \citet{Cardelli1989} extinction law.}
    \label{fig:fluxcal_spec}
\end{figure}

We also flux calibrated the spectrum of J1735 using observations of HIP 084435 (HD\,155888), a B4V star originally included in the calibration plan as a telluric standard, obtaining a single image consisting of five, 3s exposures. We followed the same reduction steps as for J1735, and corrected for telluric features using {\sc molecfit}. We used the Image Reduction and Analysis Facility \citep[{\sc iraf};][]{Tody1986} to perform the flux calibration of J1735. We first removed all hydrogen absorption lines from the telluric-corrected spectrum of HIP 084435 using {\sc iraf} task {\tt splot}, fitting a Voigt profile to the Brackett lines and subtracting the fit from the spectrum. We then calculated the sensitivity function by assuming a 17200K blackbody for the HIP 084435 spectrum, with magnitudes of $H=8.494\pm0.042$ and $K=8.459\pm0.020$ \citep{Skrutskie2006}. The resultant sensitivity function was applied to the spectrum of J1735 to derive the flux calibrated $H+K$ spectrum. We show this spectrum in Fig. \ref{fig:fluxcal_spec}, along with the dereddened spectrum, obtained using the {\sc iraf} task {\tt deredden}, assuming $E(B-V)=1.68$ \citep{Degenaar2010burst} and a \citet{Cardelli1989} extinction law, for a standard Milky Way ratio of total to selective extinction, $R_V=3.1$.

\section{Analysis and Results} 
\label{sec:results}

\subsection{Long-term nIR monitoring}

The VVV $K_s$-band differential light curve of the nIR counterpart to J1735 is shown in Fig. \ref{fig:vvv_lc}. When comparing the target/comparison differential light curves with the comparison/comparison differential light curves, we note that J1735 appears to show nIR variability with an amplitude of up to $\sim0.45$ mag. The absence of such variability in the comparison light curves implies that it is real, and intrinsic to the J1735 system. We also find that the excess variance \citep[an estimator of the intrinsic source variance, see e.g.][]{Vaughan2003} is much stronger for the target light curve than for any comparison star light curve. The average NIR magnitudes of J1735 (using the 1\arcsec\ aperture photometry from the source detection table) are $J=15.249\pm0.008$, $H=14.229\pm0.007$ and $K_s=13.694\pm0.002$.

\begin{figure}
    \centering
    \includegraphics[width=0.475\textwidth]{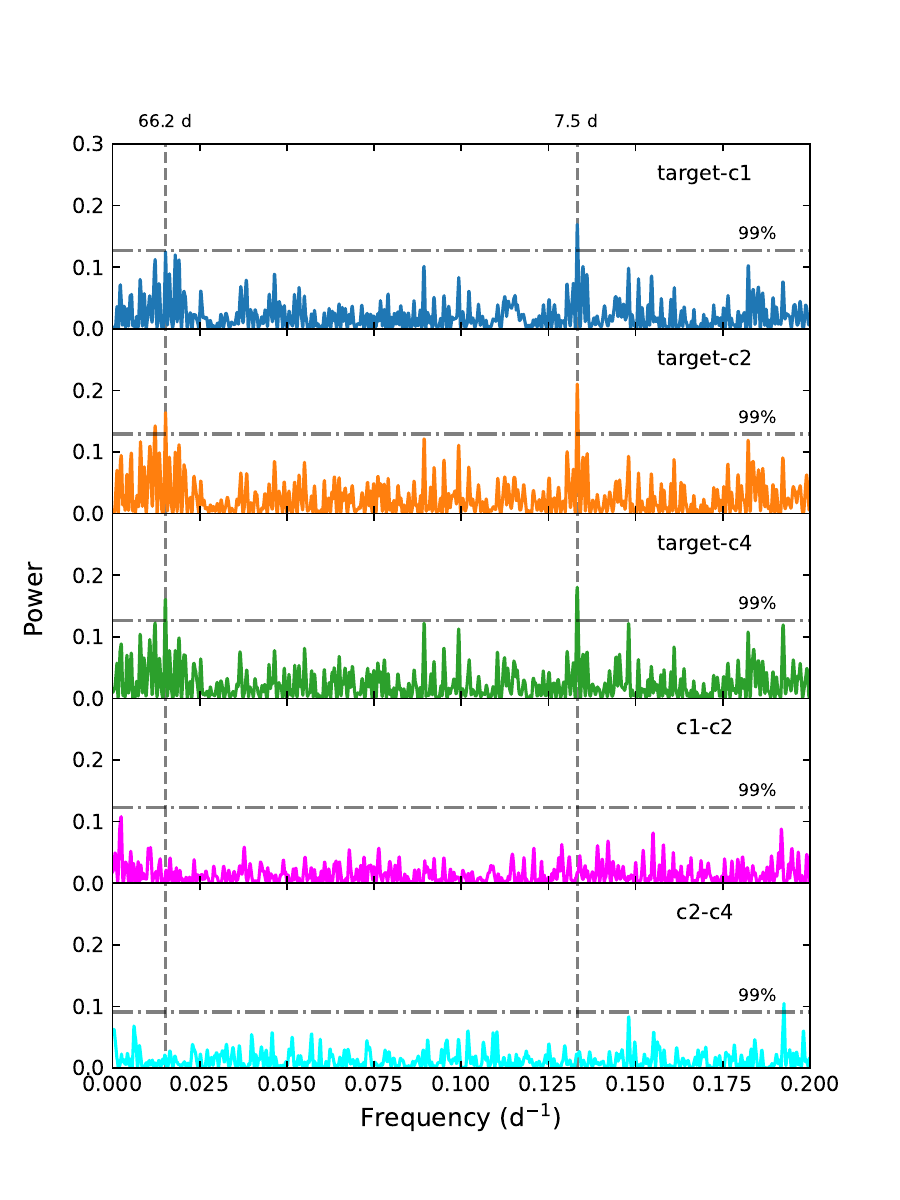}
    \caption{Lomb-Scargle periodograms of the light curves presented in Fig. \ref{fig:vvv_lc}. Two peaks are marked with dashed lines and their corresponding periodicities are labelled. The dash-dotted lines in each panel represent the 99\% significance level.}
    \label{fig:periodograms}
\end{figure}

We performed timing analysis on the differential light curves in order to search for any periodic variability that might be associated with the system. We calculated Lomb-Scargle periodograms \citep{Lomb1976,Scargle1982} for each of the five differential light curves shown in Fig. \ref{fig:vvv_lc} using the {\tt LombScargle} class included in the {\sc astropy} Python package \citep{AstropyCollaboration2013,AstropyCollaboration2018}. The periodograms are shown in Fig. \ref{fig:periodograms}. For each periodogram, we calculated a 99\% significance threshold for the power. We did this by randomly shuffling the light-curve magnitudes but keeping the time stamps the same, effectively creating a randomized light curve with the same sampling as the original data. We calculated the peak Lomb–Scargle power for 10,000 of these randomized light curves, from which we derived the 99\% significance for each periodogram.

The periodograms of the target/comparison differential light curves all show significant peaks, the two strongest of which correspond to periods of 7.5 and 66.2d, with the strongest peak (and only one to be consistently above the 99\% significance line) at $7.5\pm0.3$d. These peaks are persistent through the three target/comparison periodograms, and do not appear in the comparison/comparison periodograms, implying that the corresponding periods are real and not an artefact of the light curve sampling, though their origin is unclear.

\subsection{nIR spectroscopy}
\label{sec:nIR_spec_obs}

\begin{figure}
    \centering
    \begin{subfigure}{0.5\textwidth}
       \includegraphics[width=\textwidth]{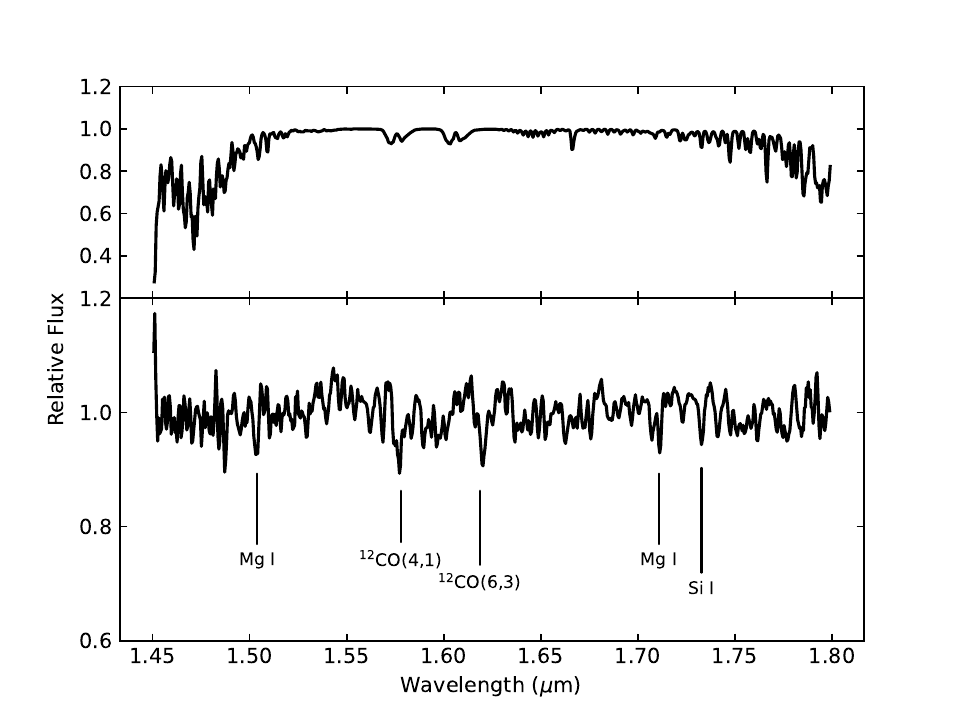}
    \end{subfigure}
    \begin{subfigure}{0.5\textwidth}
       \includegraphics[width=\textwidth]{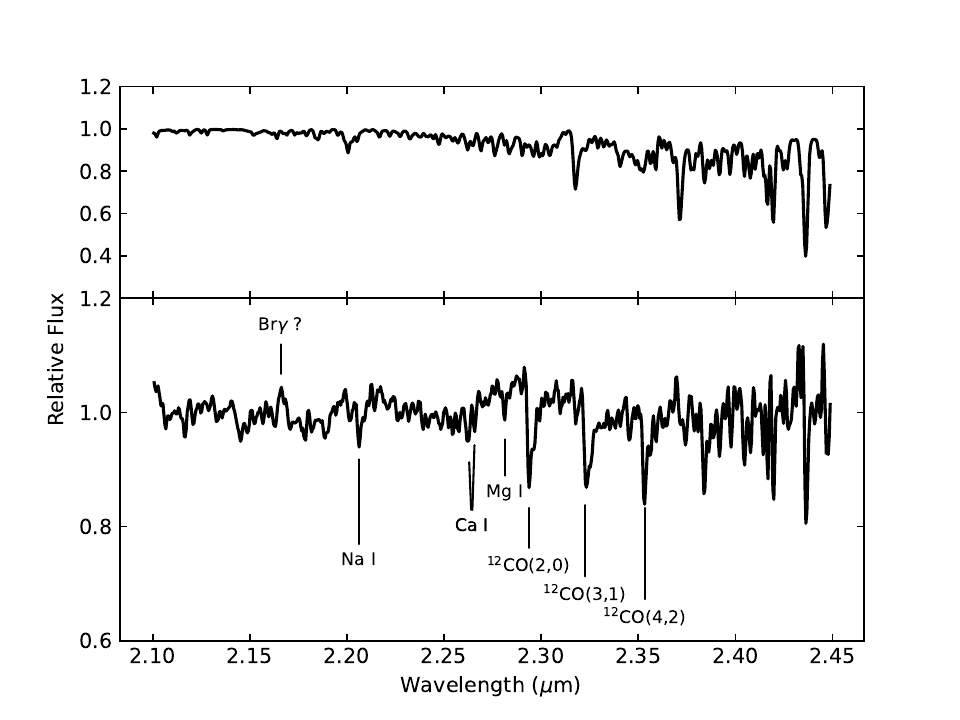}
    \end{subfigure}
    \caption{VLT/SINFONI spectrum of the nIR counterpart to J1735. In the upper sub-figure we show the $H$-band portion of the spectrum and in the lower sub-figure we show the $K$-band portion of the spectrum. In the top panel of both sub-figures we show the telluric transmission spectrum derived with {\sc molecfit} and used to correct the target spectrum. The lower panel of each sub-figure shows the continuum normalised, telluric corrected spectrum of the target. The target has been velocity corrected to the rest frame using the observed $v_{\rm obs}=45\pm11$ km s$^{-1}$ and the spectrum has been smoothed with a 3-point boxcar function. Some spectral lines have been identified and labelled.
    }
    \label{fig:spectrum}
\end{figure}

The flux-calibrated nIR spectrum of J1735 is shown in Fig. \ref{fig:fluxcal_spec}. We also show the spectrum divided into $H$- and $K$-band segments and continuum normalised in Fig. \ref{fig:spectrum}. An initial examination of the spectrum reveals numerous lines associated with neutral metals, along with molecular $^{12}$CO bandheads, classic nIR spectral signatures of late-type (K-M) stars \citep[e.g.][]{Kleinmann1986,Wallace1997,Meyer1998,Ivanov2004,Rayner2009}. Utilising the National Institute of Standards and Technology – Atomic Spectra Database \cite[NIST-ASD][]{NIST_ASD} to obtain the most accurate rest wavelengths of these features, we used the {\sc iraf} task {\tt rvidlines} task to calculate the observed radial velocity, $v_{\rm obs}$ of the source. This is a sum of the systemic and orbital radial velocity of the system, as we are unable to disentangle the two from a single spectrum. We find $v_{\rm obs}=45\pm11$ km s$^{-1}$. The spectra shown in Fig. \ref{fig:spectrum} are velocity corrected using the measured $v_{\rm obs}$.

The most striking lines in the nIR spectrum of J1735 are the strong CO bandheads at $\lambda>2.29$ $\mu$m. The presence of these bandheads (we note that CO is also present in the $H$-band), along with the relative weakness of Ca {\sc i} and Na {\sc i}, is strong evidence for a late-type giant companion \citep[see e.g.][]{Wallace1997}. To test this hypothesis, we can apply 2D spectral classification techniques to the $K$-band portion of the nIR spectrum. 
Numerous studies 
have identified temperature-dependent and surface gravity-dependent features in nIR spectra of late-type stars \citep[e.g.][]{Ramirez1997,Ivanov2004,Comeron2004,Rayner2009,Messineo2017,Ghosh2019,Messineo2021}. Comparing the strengths of these features allows one to derive the spectral type and luminosity class of the target. In the case of J1735 we assume this to be the donor star of the accreting binary. These techniques were successfully applied by \citet{Bahramian2014a} and \citet{Shaw2020} to identify and type giant and subgiant donors in a number of accreting binaries.

We adopt the Na {\sc i} (2.21$\mu$m) and Ca {\sc i} (2.26$\mu$m) passbands as the temperature dependent features \citep[e.g.][]{Ramirez1997}, and the $^{12}$CO $(2,0)$ (2.29$\mu$m) bandhead as the surface-gravity and temperature-dependent feature. For each feature we calculate the equivalent width (EW), approximating the continuum level with a 1D polynomial (fit using {\sc astropy}'s {\tt Polynomial1D} function) using two nearby, featureless regions of the spectrum. We follow the method of \citet{Shaw2020}, who define the continuum and line bandpasses according to those of \citet{Bahramian2014a} and \citet{Comeron2004} and are almost identical to the continuum and line definitions of later studies \citep[see e.g.][]{Ghosh2019}. As an additional check, we also measure the EW of the Mg {\sc i} (1.71$\mu$m) line which, when compared with $^{12}$CO $(2,0)$, provides a strong diagnostic for luminosity class \citep{Messineo2017,Messineo2021}. The line and continuum definitions, along with the measured EWs, are reported in Table \ref{tab:ews}. We also include continuum and line definitions for Si {\sc i} (1.59 $\mu$m), $^{12}$CO $(6,3)$ (1.62 $\mu$m) and Mg {\sc i} (2.28 $\mu$m), the EWs of which can be used to investigate how much, if at all, the measured spectrum is diluted by non-stellar sources \citep[such as an accretion disc;][]{ForsterSchreiber2000}. Uncertainties on the EWs were calculated using a bootstrap-with-replacement method as follows: we determined 10,000 new EWs using continuum levels from a random sample of the passbands described in Table \ref{tab:ews} and computed the standard error. For each new continuum passband, we allowed the same wavelength-flux pair to be selected multiple times. We cross-checked our EW calculations with the {\tt equivalent\_width} function provided by the {\sc astropy} package {\sc specutils} \citep{specutils} and found them to be consistent. In Table \ref{tab:ews}, we also include the integrated line fluxes of each spectral feature of interest, measured from the continuum subtracted dereddened spectrum.

Aside from the metal lines, we also note the possible presence of a Brackett $\gamma$ (Br$\gamma$) emission line at 2.17 $\mu$m (continuum and line definitions provided in Table \ref{tab:ews}). Though weak (we measure EW$=-1.64\pm0.13$ \AA), it is consistent with the presence of hydrogen emission in the optical spectrum \citep{Degenaar2010burst}. We see no further evidence of hydrogen emission in the nIR spectrum of J1735.

\begin{table*}
    \centering
    \caption{Measured EWs and fluxes of spectral features of the nIR spectrum of J1735. We also define the locations of the features and the passbands used to estimate the continuum,  references for which are provided in the last column of the table}.
    \begin{tabular}{lccccccccc}
    \hline
    & \multicolumn{2}{c}{Line} & \multicolumn{2}{c}{Blue Continuum} & \multicolumn{2}{c}{Red Continuum} & & & \\
    Feature & Centre  & $\Delta\lambda$  & Centre & $\Delta\lambda$ & Centre & $\Delta\lambda$ & EW  & Flux & Reference\\
    & ($\mu$m) & ($\mu$m) & ($\mu$m) & ($\mu$m) & ($\mu$m) & ($\mu$m) & (\AA) & (erg cm$^{-2}$ s$^{-1}$) & \\
    \hline\hline
    Si {\sc i} & 1.5890 & 0.004 & 1.5859 & 0.0015 & 1.5937 & 0.0015 & $1.03\pm0.13$ & $-2.5\times10^{-15}$ & (1)\\
    $^{12}$CO $(6,3)$ & 1.6198 & 0.0045 & 1.6160 & 0.003 & 1.6270 & 0.003 & $2.63\pm0.24$ & $-5.8\times10^{-15}$ & (2) \\
    Mg {\sc i} & 1.7115 & 0.005 & 1.7035 & 0.001 & 1.7145 & 0.002 & $1.81\pm0.14$ & $-3.3\times10^{-15}$ &(1)\\
    \hline
    Br$\gamma$ & 2.1661 & 0.009 & 2.1560 & 0.012 & 2.1745 & 0.005 & $-1.64\pm0.13$ & $1.0\times10^{-15}$ & (3)\\
    Na {\sc i} & 2.2075 & 0.007 & 2.1940 & 0.006 & 2.2150 & 0.004 & $2.13\pm0.19$ & $-1.3\times10^{-15}$ & (4)\\
    Ca {\sc i} & 2.2635 & 0.011 & 2.2507 & 0.0106 & 2.2710 & 0.002 & $1.18\pm0.24$ & $-5.9\times10^{-16}$ & (4)\\
    Mg {\sc i} & 2.2807 & 0.0035 & 2.277 & 0.0040 & 12.2865 & 0.008 & $0.85\pm0.06$ & $-4.3\times10^{-16}$ & (4) \\
    $^{12}$CO $(2,0)$ & 2.2955 & 0.013 & 2.2500 & 0.016 & 2.2875 & 0.007 & $8.66\pm0.35$ & $-4.2\times10^{-15}$ & (4) \\
    \hline
    \end{tabular}\\
    References for line and continuum passbands: (1) \citet{Messineo2017}, (2) \citet{Ghosh2019}, (3) This work, (4) \citet{Comeron2004}
    \label{tab:ews}
\end{table*}

\section{Discussion}

\subsection{Spectral type of the companion}

\subsubsection{Luminosity class}

Table \ref{tab:ews} shows the EWs of lines important for deriving the spectral type of the donor star in J1735. The first thing to note is that we measure EW[$^{12}$CO (2,0)]$=8.66\pm0.35$ \AA. \citet{Comeron2004} note that supergiants typically exhibit EW[$^{12}$CO (2,0)]$\gtrsim25$ \AA, while all the supergiants in the Infrared Telescope Facility (IRTF) spectral library of cool stars \citep{Rayner2009} show EW[$^{12}$CO (2,0)]$\gtrsim20$ \AA\ \citep{Messineo2017,Messineo2021}. We are therefore able to confidently rule out a supergiant donor for J1735, implying that it is instead a dwarf, giant, or subgiant.

We can compare the EWs of $^{12}$CO $(2,0)$ to those of Na {\sc i} and Ca {\sc i} in order to determine the luminosity class of the donor star. \citet{Ramirez1997} showed that dwarfs exhibit $-0.22<\log_{10}$\{EW[$^{12}$CO (2,0)]/(EW[Na {\sc i}]$+$EW[Ca {\sc i}])\}$<0.06$ while giants show much higher ratios in the range $0.37<\log_{10}$\{EW[$^{12}$CO (2,0)]/(EW[Na {\sc i}]$+$EW[Ca {\sc i}])\}$<0.61$. \citet{Villaume2017} also note that EW[$^{12}$CO (2,0)] alone is a strong discriminator between giant and dwarf stars for all metallicities. 
We measure $\log_{10}$\{EW[$^{12}$CO (2,0)]/(EW[Na {\sc i}]$+$EW[Ca {\sc i}])\}$=0.42\pm0.04$, strongly indicative of a giant companion. Another piece of evidence for a giant classification comes from EW[Mg {\sc i} (1.71$\mu$m)], which, when compared with EW[$^{12}$CO (2,0)], is consistent with known giants \citep[see figure 3 of][]{Messineo2021}. The strength of the Mg {\sc i} (1.71$\mu$m) line alone is also a good indicator of a giant classification, as late-type dwarfs typically show EW[Mg {\sc i} (1.71$\mu$m)]$>2$ \citep{Rayner2009}. We can therefore confidently identify the companion star of J1735 as a late-type giant. 

\subsubsection{Contamination of the donor star spectrum by the accretion disc}

We can also use EW ratios of temperature-dependent features to assess whether there is a source of diluting continuum emission that would affect the conclusions we draw from the measured EWs. Given that there is evidence for Br$\gamma$ emission in the $K$-band spectrum, it is possible that there could be continuum emission from the accretion disc contributing the nIR spectrum. \citet{ForsterSchreiber2000} demonstrated that ratios of temperature-dependent lines close in wavelength can provide an indication of continuum dilution in the $H$- and $K$-bands. In their figure 8, \citet{ForsterSchreiber2000} show plots of dilution-free EW ratios, with arrows indicating how measured ratios would change with a source of dilution. For J1735, our measured values for EW[$^{12}$CO (2,0)], EW[Ca {\sc i}] and EW[Mg {\sc i} (2.28$\mu$m)] strongly suggest that there is negligible dilution in the $K$-band, as our measured ratios all lie on the loci for non-diluted sources. However, the ratio of the $^{12}$CO (6,3) and Si {\sc i} lines in the $H$-band lies below the observed distribution of non-diluted stars presented by \citet{ForsterSchreiber2000}, implying that there could be some contamination in this portion of the nIR spectrum. We therefore focus on the $K$-band spectrum when deriving the effective temperature of the donor star.


\subsubsection{Effective temperature}
\label{sec:Teff}

We can use the measured EWs (typically of EW[$^{12}$CO (2,0)]) to also derive the effective temperature, $T_{\rm eff}$, and therefore the spectral subtype of the donor star. However, metallicity can play a role in the relationship between EW and $T_{\rm eff}$. \citet{Ghosh2021} derive a relationship between $x=$EW[Na {\sc i}] and metallicity $z=$[Fe/H] based on $R\sim1200$ $K$-band spectra of 260 late-type giants:


\begin{equation}
   z = m0+ax + cx^2 +ex^3, 
   \label{eq:ghosh_metal}
\end{equation}

where $m0=-2.328\pm0.080$, $a=1.863\pm0.109$, $c=-0.485\pm0.042$ and $e=0.045\pm0.005$. Though the standard error of estimate (SEE) on the relation is large (SEE $\sim0.25$ dex), it provides a first estimate of [Fe/H]. Assuming EW[Na {\sc i}]=$2.13\pm0.19$, we find [Fe/H]$=-0.1\pm0.4$, propagating uncertainties and factoring in the SEE. Despite the large scatter, the EW of the Na {\sc i} line is indicative of a donor that is not unusually metal-rich or metal-poor.

\citet{Ghosh2021} also derive a linear relationship between $T_{\rm eff}$ and EW[$^{12}$CO]:

\begin{equation}
   z = m0+ax, 
   \label{eq:ghosh_temp}
\end{equation}

where $z$ and $x$ in this instance refer to $T_{\rm eff}$ and EW[$^{12}$CO (2,0)], respectively. For the full sample, \citet{Ghosh2021} derive $m0=5370\pm30$ and $a=-82.7\pm2.0$, from which we infer $T_{\rm eff}=4650\pm160$, factoring in the large scatter (SEE $=149$). If, instead, we assume $-0.3\leq{\rm [Fe/H]}\leq0.3$, which is likely given our calculation of [Fe/H]$=-0.1\pm0.4$, then \citet{Ghosh2021} derive $m0=5651\pm44$ and $a=-99\pm3$, which would increase the temperature to $T_{\rm eff}=4790\pm140$. Both ranges of $T_{\rm eff}$ are indicative of a spectral type in the range G6III -- K1III according to the $T_{\rm eff}$ -- spectral type relations of \citet{vanBelle2021}, putting the radius of the companion in the range $9R_{\rm \odot}\lesssim R_{\rm c}\lesssim16R_{\rm \odot}$ \citep[see table 17 of][]{vanBelle2021}. 

Deriving the spectral type of the companion also allows us to further investigate whether the Br$\gamma$ feature described in Section \ref{sec:nIR_spec_obs} is real or chance coincidence. All giants in the IRTF spectral library of cool stars in the range G6III -- K1III show weak, but detectable, Br$\gamma$ absorption in the $K$-band. Subtracting continuum-normalised IRTF standard spectra from our (continuum-normalised) SINFONI spectrum of J1735 therefore highlights the apparent Br$\gamma$ emission in the source spectrum even further. An F-test of the Br$\gamma$ region of the spectrum of J1735, comparing the spectrum with and without a best-fit Gaussian line at 2.166$\mu$m indicates that the line is significant above 99.9\% confidence for any  IRTF standard in the G6III -- K1III spectral type range, and thus we believe it to be real.

\subsubsection{Surface gravity and mass}
\label{sec:mass}

\citet{Ghosh2019} derived a linear relationship between the surface gravity ($\log g$) and EW[$^{12}$CO (2,0)] with the same form as Equation \ref{eq:ghosh_temp}, where $z$ is now $\log g$ and $x$ is EW[$^{12}$CO (2,0)]. Using the best-fit values $m0=3.75\pm0.13$ and $a=-0.16\pm0.01$ we derive $\log g = 2.4\pm0.35$ cm s$^{-2}$ (factoring in SEE $=0.29$). Though the uncertainty is relatively large, we can use this value to place a rough estimate on the mass of the donor star. Taking a median $R_{\rm c}\sim12.5\Msun$ (see Section \ref{sec:Teff}, above) and $\log g \sim 2.4$ cm s$^{-2}$ we estimate that the mass of the companion is $M_{\rm c}\sim1.4\Msun$ - similar to the NS primary. We note that this is not a detailed calculation of $M_{\rm c}$ as \citet{Ghosh2019} calibrated their relation using only K and M giants, and we have established that the range of $T_{\rm eff}$ we have derived for the donor of J1735 can include G stars. In addition, the large range of valid stellar radii and $\log g$ mean this number is also subject to a significant scatter. However, $M_{\rm c}\sim1.4\Msun$ remains a useful estimate when we consider the nature of accretion in the J1735 system. 

\subsection{Timing analysis}
\label{sec:timing}

\begin{figure}
    \centering
    \includegraphics[width=0.475\textwidth]{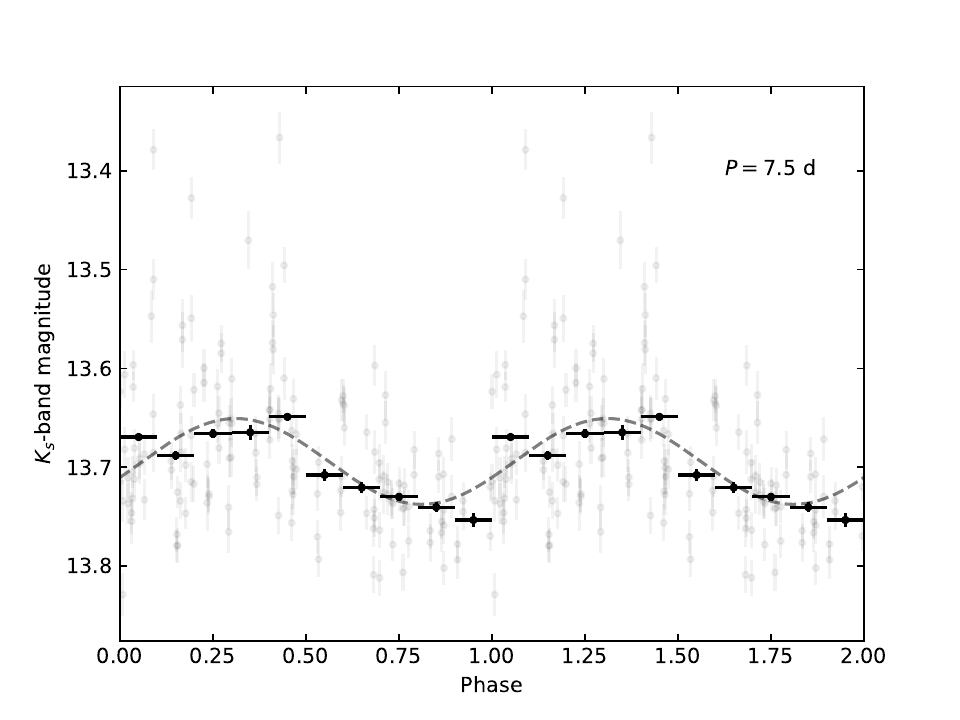}
    \caption{The VVV $K_s$-band light curve of the nIR counterpart to J1735, folded on the strongest period from the timing analysis of $7.5$d. The grey points represent the unbinned folded light curve; black points are grouped into 10 bins per cycle. The dashed curve represents the best-fit sinusoidal model as determined by the Lomb-Scargle method.}
    \label{fig:7p5d_fold}
\end{figure}

The periodograms presented in Fig. \ref{fig:periodograms} imply at least two statistically significant periodicities associated with the J1735 system, the strongest of which is at $7.5\pm0.3$d. In Fig. \ref{fig:7p5d_fold} we show the $K_s$-band light curve of J1735, folded on this periodicity, along with the best-fit sinusoidal model as determined by the {\sc astropy} {\tt LombScargle} method. Fig. \ref{fig:7p5d_fold} shows that the phase-folded light curve follows a sinusoidal pattern with significant scatter about the mean, possibly indicative of short-term flaring episodes. Despite the high significance of the $P=7.5$d peak in the periodogram (a false alarm probability of FAP=$0.02$\%), the folded light curve does not provide any insight into the nature of the variability. If $P=7.5$d is the orbital period of the binary, then a future nIR spectroscopic radial velocity campaign will be able to confirm this.

\subsection{The nature of 1RXH J173523.7-354013: wind accretion or Roche-lobe overflow?}
Having identified the donor star in J1735 as a late-type giant through nIR spectroscopy, we can now turn to the X-ray properties of the system to discuss the binary configuration. Any plausible scenario must be able to explain i) its low persistent X-ray luminosity of $\sim 10^{34}-10^{35}~\lum$ (see Section \ref{sec:dist}), ii) its ability to ignite thermonuclear bursts, and iii) the large optical increase observed during its thermonuclear burst detection \citep[][]{Degenaar2010burst}. It has already been established that J1735 cannot be a UCXB, owing to the detection of H$\alpha$ in its optical spectrum. This is further confirmed by the detection of molecular lines in the nIR spectrum that indicate a giant companion. We therefore explore two more potential scenarios here: Scenario 1 in which J1735 is an LMXB, accreting via Roche lobe overflow (RLOF) and Scenario 2, in which J1735 is a wind-fed symbiotic X-ray binary.

\subsubsection{A note on the distance}
\label{sec:dist}

\citet{Degenaar2010burst} calculate an upper limit on the distance to J1735 to be $d\lesssim9.5$ kpc, inferred from the peak flux of the X-ray burst. However, \citet{Degenaar2010burst} note that the peak flux was only measured in the hard X-ray band, and extrapolated to lower energies, hence a large upper limit. \citet{BailerJones2021}, on the other hand, calculated the photogeometric distance to J1735 to be $d=3.5^{+2.3}_{-1.3}$ kpc, from {\em Gaia} EDR3.

\citet{Degenaar2010burst} calculate a persistent unabsorbed X-ray flux of $F_{\rm X}\sim1.9\times10^{-11}$ erg cm$^{-2}$ s$^{-1}$ in the 0.5--10 keV band, and a persistent {\em bolometric} flux of $F_{\rm bol}\sim3.8\times10^{-11}$ erg cm$^{-2}$ s$^{-1}$ (0.01--100 keV). Combining these measured fluxes with the \citet{BailerJones2021} distance range implies that J1735 has a persistent X-ray luminosity in the range $1.1\times10^{34}\lesssim L_{\rm X}\lesssim 7.7\times10^{34}$ erg s$^{-1}$, and bolometric luminosity $2.2\times10^{34}\lesssim L_{\rm bol}\lesssim 1.5\times10^{35}$ erg s$^{-1}$. 

These values are much lower than the upper limits calculated by \citet{Degenaar2010burst}, which implies that the global mass-accretion rate on to the NS, $\dot{M}$, is also lower. For a NS of mass and radius $M_{\rm NS}=1.4\Msun$ and $R_{\rm NS}=10$ km, respectively, this would imply $1.9\times10^{-12}\lesssim \dot{M} \lesssim 1.3\times10^{-11}\Msun$ yr$^{-1}$.

\subsubsection{Scenario 1: Roche Lobe Overflow}
\label{sec:RLOF}

The first possibility that we must consider is that J1735 is an LMXB undergoing RLOF, but with a giant companion rather than the typical main sequence secondary. If we assume that the periodicity measured in Section \ref{sec:timing} is the orbital period $P_{\rm orb}$ of the binary and that the binary components are the same mass ($M_{\rm NS} = M_{\rm c} \sim 1.4\Msun$; Section \ref{sec:mass}) then we can estimate the orbital separation, $a\sim1.6\times10^{12}$ cm, using Kepler's Third Law. The approximate radius of the Roche lobe in a binary system with equal masses is $R_{\rm RL} = 0.38a$ \citep{Eggleton1983}. Utilizing the same assumption above regarding $P_{\rm orb}$, as well as $q\sim1$ (where $q=M_{\rm c}/M_{\rm NS}$ is the binary mass ratio), we calculate $R_{\rm RL}\sim6.1\times10^{11}$ cm, or $\sim9\Rsun$.

Given the range of stellar radii given by \citet{vanBelle2021} for a G6III -- K1III star, it seems plausible that the companion would exceed its Roche Lobe. However, this is heavily dependent on the assumption that $P_{\rm orb}=7.5$ d. If the $P_{\rm orb}$ of J1735 is instead 66.2 d, where a peak is also seen in some of the periodograms in Fig. \ref{fig:periodograms}, a late G/early K giant would not fill the Roche Lobe. In addition, any change in component masses that would result in the assumed $q\sim1$ being higher, such as a more massive companion, or lower mass NS, increases the size of the Roche lobe, such that the lower end of the $R_{\rm c}$ range calculated in Section \ref{sec:Teff} would not fill it in the $P_{\rm orb}=7.5$ d configuration.\footnote{We note that the opposite can also be true, such that a lower value of $q$ could reduce the size of the Roche Lobe.} 

If $P_{\rm orb}=7.5$ d and the system is indeed a RLOF system, then $\dot{M}_{\rm RLO}\sim10^{-8}\Msun$ yr$^{-1}$ \citep{Webbink1983}, and thus mass transfer would have to be highly non-conservative to match the calculated $\dot{M}$ on to the NS. For several NS-LMXBs in compact orbits it has been proposed that mass-transfer is highly non-conservative based on timing studies of the X-ray pulsations \citep[][]{disalvo2008,Burderi2009,Marino2019}, or by modeling the multi-wavelength spectral energy distribution of the accretion disc \citep[][]{Hernandez2019}. However, if severe mass losses are indeed the explanation for these observations, it is unclear what should be causing this and if the same could be happening in J1735. Severe mass losses, in principle, can be observed through future dedicated multi-wavelength studies of the system, e.g. as an additional source of absorption in X-rays \citep[see e.g.][]{Iaria2021}, or as the origin of reprocessed emission in the mid -- far IR (which would require JWST).

One possibility worth considering then is if J1735 could be a RLOF system `in quiescence', in the sense that its accretion disk is currently cold and neutral but steadily filling up until sufficient mass is accumulated for it to become ionized and an outburst to occur \citep[see][for a recent review on this disk instability paradigm]{hameury2020}. There is an empirical relation between the orbital period and quiescent X-ray luminosity of (BH) LMXBs accreting via RLOF (\citealt{garcia2001}; see also \citealt{Carotenuto2022} for a recent representation), where a longer $P_{\rm orb}$ is expected to lead to a higher quiescent $L_{\rm X}$ due to the existence of a larger disk. In the work of \citet{garcia2001}, there are no NSs with known orbital periods on the order of $\sim 100$~hr and measured quiescent luminosities, but there are a few transient BH-LMXBs with long periods that have $L_{\rm X} \sim 10^{33}-10^{34}~\lum$. At the same time, we know that for similar orbital periods, NS systems typically have a factor $\sim10-100$ higher quiescent luminosities than the BH ones \citep[due to NSs having a solid surface;][]{Menou1999}. While speculative, it does not seem implausible that J1735 is a quiescent LMXB with $L_{\rm X} \sim 10^{34}-10^{35}~\lum$. One might then expect such a system to spend a long time in this quiescent state, filling up the large disk, before launching into a prolonged outburst. If true, such an outburst might further be expected to be very bright, reaching mass-accretion rates around the Eddington limit, as is seen in the NS Z-sources \citep[e.g.,][]{Hasinger1989,Homan2010}  It is worth noting that the calculated $\dot{M}$ lies well inside the unstable (transient) regime predicted by the disk-instability model that reproduces many of the features of LMXB outbursts, assuming $P_{\rm orb}=7.5$ d \citep{Lasota2001,Coriat2012}.

We briefly note that the X-ray spectral properties of J1735 do not exclude the luminous quiescence scenario. An \xmm\ observation of J1735 showed that it consists of a $\simeq0.3$~keV black-body component, plausibly emerging from the NS surface, and a $\Gamma \simeq 1.4$ power law \citep[][]{ArmasPadilla2013}. This is consistent with many other transient NS-LMXBs that have lower X-ray luminosities in quiescence \cite[e.g.,][]{Rutledge2001,Wijnands2002a,Campana2003,Jonker2003}. However, it is also true that short-period NS-LMXBs accreting at low rates have similar X-ray spectra \citep[e.g.,][]{degenaar2017,ArmasPadilla2013,Papitto2013b}.

Following \citet{vanParadijs1994}, we can also test whether the observed burst properties work within the framework of optical reprocessing of the X-ray emission by an accretion disk. The peak of the X-ray burst was a factor $\sim1000$ brighter than the persistent X-ray flux \citep{Degenaar2010burst}. We would therefore expect an increase in optical flux by $\sim3.4$ mag, provided it is dominated by an accretion disk \citep{vanParadijs1994}. \citet{Degenaar2010burst} measured an optical increase of 3.2 mag, well within the scatter of the \citet{vanParadijs1994} relation. The observed increase in optical brightness is therefore consistent with the reprocessing of the X-ray burst in the accretion disk that one might expect in a RLOF LMXB.

Finally, we briefly consider the possibility that J1735 is a RLOF LMXB that is not X-ray faint because it is accreting at very low rate, but due to a very high inclination prohibiting us to see the bright central X-ray source. A number of such LMXBs are known and are referred to as Accretion Disk Coronae (ADC) sources because their X-ray emission comes from a small fraction of the central source
that is scattered in material around the disk \cite[e.g.,][]{White1982,Ryota2023}. If J1735 is an ADC source, the mass-accretion rate inferred from the observed X-ray luminosity is highly underestimated, and that it could in fact be as high as that expected for a 7.5-d binary. However, the multi-wavelength properties of J1735 do not show indications of a high inclination (such as a high X-ray absorption, dipping X-ray variability, or a flat optical/infrared spectral energy distribution). Moreover, in this scenario it would be difficult to explain why the system significantly brightened during its burst, by a factor of $\sim 1000$ in X-rays and by $\sim 3$~mag in the UV/optical band \citep[][]{Degenaar2010burst}. Finally, we note that the long duration of that burst is consistent with that seen in other low-$\dot{M}$ NS-LMXBs \citep[e.g.,][]{Falanga2009,Bozzo2015,degenaar2011burst,degenaar2013c}. The ADC scenario therefore does not seem likely for J1735.

\subsubsection{Scenario 2: A wind-fed system}
\label{sec:wind}

Given the propensity for giant stars to drive dense stellar winds, it is worth investigating if the companion of J1735 is able to drive a strong enough wind to produce the observed X-ray emission. \citet{Schroeder2005} developed a semi-empirical formula for the mass loss of cool stellar winds, building upon \citet{Reimers1975} law. The mass loss rate, $\dot{M}_{\rm w}$, according to this relation, can be described as such:

\begin{equation}
    \dot{M}_{\rm w} = \eta \frac{L_*R_*}{M_*} \left(\frac{T_{\rm eff}}{4000 {\rm K}}\right)^{3.5} \left( 1 + \frac{g_{\odot}}{4300g_*}\right)
\end{equation}

\noindent where $L_*$, $R_*$ and $M_*$ are the star's luminosity, radius and mass in solar units, respectively, $g_*$ and $g_{\odot}$ and are the star's surface gravity and solar surface gravity, respectively and $\eta$ is a fitting parameter given as $\eta\sim8\times10^{-14}\Msun$ yr$^{-1}$.

As we have already established that a $12.5\Rsun$ star would fill its Roche Lobe, we can only investigate the wind scenario if we assume that the radius of the companion star is on the lower end ($R_{\rm c}\sim9\Rsun$) of the range given in Section \ref{sec:Teff} \citep{vanBelle2021}. Making a simple assumption on the luminosity of the companion based from the Stefan-Boltzmann law, and assuming the same values for mass and surface gravity as in Section \ref{sec:mass}, we derive $L_*\sim 40 {\rm L}_{\odot}$ and thus $\dot{M}_{\rm w}\sim4\times10^{-11} \Msun$ yr$^{-1}$.

By adopting a Bondi-Hoyle-Lyttleton formalism \citep{Hoyle1941,Bondi1944}, we can estimate the mass accretion rate on to the NS in the wind-accreting scenario:

\begin{equation}
    \dot{M} = \frac{3}{16} \left (\frac{R_{\rm c}}{a} \right )^2 \frac{q^2}{\beta_{\rm w}^4} \left ( 1 + \frac{1+q}{2\beta_{\rm w}^2}\frac{R_{\rm c}}{a}\right )^{-\frac{3}{2}} |\dot{M}_{\rm w}|.
\end{equation}

\noindent Where $a$ is the orbital separation, $q$ is the binary mass ratio ($M_{\rm NS}/M_{\rm c}$ in this case, as opposed to the inverse in Section \ref{sec:RLOF}), and $\beta_{\rm w}$ is the wind velocity parameterized as a fraction of the companion star's escape velocity \citep[see][]{Willems2003}. As discussed above, if the 7.5 d periodicity measured in Section \ref{sec:timing} is $P_{\rm orb}$, then $a\sim1.6\times10^{12}$ cm, assuming the mass of each star is $1.4\Msun$. 

Using $q=1$, as well as $R_{\rm c}\sim9$ $\Msun$ and $\dot{M}_{\rm w}\sim4\times10^{-11} \Msun$ yr$^{-1}$, we estimate $\dot{M}\sim7\times10^{-13} \Msun$ yr$^{-1}$, assuming that the wind velocity is equivalent to the escape velocity of the companion ($\beta_{\rm w}=1$). This is inconsistent with the $\dot{M}$ calculated in Section \ref{sec:dist}. Based on our assumptions, we find that Bondi-Hoyle-Lyttleton accretion is highly unlikely to be driving the observed properties of J1735.

Considering that a 9$\Rsun$ star may not quite fill the Roche Lobe, especially if some of the assumptions regarding the mass ratio change even slightly, it is possible that its stellar wind is gravitationally focused in the direction of the NS \citep{Friend1982,deVal-Borro2009}. \citet{deVal-Borro2017} calculate that the mass-accretion rate on to the compact object can be up to 5--20\% of the total wind mass-loss rate from the donor in the focused wind case. For J1735, this would imply that $\dot{M}$ can be as high as $8\times10^{-12}$ $\Msun$ yr$^{-1}$, adopting the derived $\dot{M_{\rm w}}$ for a $9\Rsun$ companion above. This is well within the calculated range of $\dot{M}$ in Section \ref{sec:dist}. We do emphasize that this calculation depends heavily on many assumptions about the binary components, and this scenario only works if the companion star is a relatively small giant.

\section{Conclusions}

In this work, we have examined the $H$- and $K$-band spectrum of the nIR counterpart to the VFXB 1RXH\,J173523.7$-$354013. We find that the $K$-band spectrum is dominated by the companion star, which we derive to be a late G or early K giant. This makes J1735 the second NS system originally identified as a VFXB now known to have a giant companion, after IGR\,J17445$-$2747 \citep{Shaw2020}. Timing analysis of nIR photometry of J1735 has revealed a strong modulation on a period of $7.5\pm0.3$d, which we postulate to be the orbital period of the system. Other periods are present in the periodograms, but only the $7.5$d period is consistently detected above 99\% confidence, regardless of the choice of comparison star.

Given the most likely properties of the binary, we consider the most likely accretion mode for J1735 to involve Roche lobe overflow, rather than accretion from the giant companion's wind. However, we cannot rule out accretion from a focused wind. RLOF requires a large mass-transfer rate ($\dot{M}_{\rm RLO}\sim10^{-8}$ $\Msun$ yr$^{-1}$) and thus mass-transfer on to the NS would have to be highly non-conservative in order to be consistent with the calculated $\dot{M}$ on to the NS (Section \ref{sec:dist}). Alternatively, we hypothesize that J1735 could be a transient system that is currently in a state in which its disk is building up to an outburst. The fact that it is much brighter than typical LMXBs in such a `quiescent' state could be explained by its large accretion disk. This scenario predicts that the source at one point would exhibit a very bright and long outburst, appearing like a Z source. 

In order to solidify our conclusions surrounding J1735, the next step will be to confirm $P_{\rm orb}$ with a nIR radial velocity campaign, which will not only confirm the orbital period, but will also allow us to derive the mass function of the binary, providing more constraints on the true mass of the companion. If the source is indeed the RLOF LMXB we claim it to be, then we may also be able to detect ellipsoidal modulations in time-series photometry, which will allow us to model the inclination of the binary.

Regardless of the true nature of accretion in J1735, it is becoming clear that compact objects with evolved companions are emerging as a significant contributor to the growing population of faint X-ray sources. \citet{Shaw2020} uncovered one symbiotic X-ray binary (IGR\,J17445$-$2747), as well as a cataclysmic variable with a subgiant donor (3XMM\,J174417.2$-$293944) in their follow-up of sources discovered in a systematic X-ray survey of the Galactic bulge \citep{Bahramian2021}. 

\section{Acknowledgments}
We thank the anonymous referee for their insightful comments that helped improve this manuscript. AWS thanks the hospitality of the Anton Pannekoek Institute, where part of this work was carried out. COH is supported by NSERC Discovery Grant RGPIN-2016-04602.

Based on observations collected at the European Organisation for Astronomical Research in the Southern Hemisphere under ESO programme 0102.D-0650(C). This research made use of {\sc astropy} ({\href{https://www.astropy.org}{https://www.astropy.org}), a community-developed core {\sc python} package for Astronomy \citep{AstropyCollaboration2013,AstropyCollaboration2018}.

\section*{Data Availability}
The SINFONI observations of 1RXH J173523.7$-$354013 are publicly available in the European Southern Observatory Science Archive Facility, along with the corresponding raw and processed calibration data. The VVV data used in this work are publicly available from the VISTA Science Archive.

\bibliographystyle{mnras}
\bibliography{references.bib}



\bsp	
\label{lastpage}
\end{document}